\documentstyle[prb,aps,epsfig]{revtex}

\begin{document}
\draft
%
%%%%%%%%% 
% Title %
%%%%%%%%%
\title{Ten dimensional Wavepacket Simulations of Methane Scattering.}
\author{R. Milot and A.P.J. Jansen}
\address{Schuit Institute of Catalysis, T/TAK, Eindhoven University of
  Technology\\ P.O. Box 513, NL-5600 MB Eindhoven, The Netherlands}
\date{\today}
\maketitle

%
%%%%%%%%%%%%
% Abstract %
%%%%%%%%%%%%
\begin{abstract}
We present results of wavepacket simulations of scattering of an
oriented methane molecule from a flat surface including all nine
internal vibrations. At a translational energy up to 96 kJ/mol we
find that the scattering is almost completely elastic. Vibrational
excitations when the molecule hits the surface and the corresponding
deformation depend on generic features of the potential energy
surface. In particular, our simulation indicate that for methane to
dissociate the interaction of the molecule with the surface should lead
to an elongated equilibrium C--H bond length close to the surface.
\end{abstract}

%
%%%%%%%%%%%%%%%%
% INTRODUCTION %
%%%%%%%%%%%%%%%%
\section{Introduction}

The dissociation of methane on transition metals is an important
reaction in catalysis. It is the rate limiting step in steam reforming
to produce syngas.\cite{hoo80} It is also prototypical for C--H
activation in other processes.  A large number of molecular beam
experiments in which the dissociation energy was measured as a function
of translational
energy\cite{ret85,ret86,lee86,lee87,cey87,bec89,hol95,hol96,lun89,xxx11,xxx13,val96a,val96b,ver93,ver94,ver95,see97}
as well as various bulb gas
experiments\cite{bee87,cho90,han91,cam93,lun94,olg95} have already been
done on this system. These experiments have contributed much to our
understanding of the mechanism of the dissociation. However, our
knowledge is still small compared to what we know about a well-studied
reaction as ${\rm H}_2$ dissociation.

Some years ago we performed a multi-configurational time-dependent
Hartree (MCTDH) study of CH${}_4$ dissociation on Ni(111)\cite{jan95}
with a potential energy surface (PES) based on our earlier
density functional theory (DFT)
calculations.\cite{bur93a,bur93b,bur93c,bur94,bur95a} We included the
distance of the methane molecule to the surface, a C--H distance, and
the orientation of methane as coordinates. Other wavepacket simulations
focused on (some of) these coordinates in combination with lattice
motion on several metals.\cite{lun91,lun92,lun95,car98} Other
theoretical studies on this system used DFT
calculations\cite{kra96,yan92,liao97} too and classical stochastic trajectory
simulations.\cite{ukr94,sti96}

None of the wavepacket simulations published so far have looked at the role
of the internal vibrations of methane. It was observed experimentally
that vibrationally hot ${\rm CH}_4$ dissociates more readily than cold
${\rm CH}_4$, with the energy in the internal vibrations being about as
effective as the translational energy in inducing
dissociation.\cite{ret85,ret86,lee87,hol95,lun89} However, a more
detailed assessment of the importance of the internal vibrations could
not be made, because of the large number of internal vibrations.  A
recent DFT calculation also showed that the transition state for
CH${}_4$ dissociation on Ni(111) involves considerable internal
excitation of the molecule.\cite{kra96} In this paper we report on
wavepacket simulations that we have done to determine which and to what
extent internal vibrations are important for the dissociation. We are
not able yet to simulate the dissociation including all internal
vibrations. Instead we have simulated the scattering of methane, for
which all internal vibrations can be included. By looking at vibrational
excitations and the deformation of the molecule when it hits the
surface we can derive information that is relevant for the dissociation.
We have used PESs that have been developed with
Ni(111) in mind, but our results should hold for other surfaces as well.

We have again used the MCTDH method for the wavepacket simulations, because
it can deal with a large number of degrees of freedom and with large
grids.\cite{man92,jan93} This method has been applied successfully to
gas phase reactions and reactions at
surfaces.\cite{man92b,man93,ham94,liu95,fan95b,fan94,fan95,cap95,wor96,jac96,ger97,bec97}

The rest of this paper is organized as follows. We start with a brief
description of the MCTDH method. Then the various PESs that we have used
are derived. (A harmonic intramolecular PES is adapted to include
anharmonicities in the C--H distance, the decrease of the C--H bond
energy due to interactions with the surface, and the increase of the
C--H bond length also due to interactions with the surface.) The results
of the simulations are presented and discussed next. We focus on
excitation probabilities and deformation of the molecule when it hits
the surface. The implications for the dissociation are discussed
separately. We end with a summary and some general conclusions.

%
%%%%%%%%%%%%%%%%%%%%%%%%%
% Computational details %
%%%%%%%%%%%%%%%%%%%%%%%%%
\section{Computational details}
%%% MCTDH METHOD %%%
\subsection{The MCTDH Method}
We give here a short overview of the MCTDH approximation for
completeness. More details can be found in Refs.~\onlinecite{man92} and
\onlinecite{jan93}. The 
exact wave-function of a $D$-dimensional system, is approximated by an
expression of the form
\begin{equation}
\label{eD}
  \Psi_{\rm MCTDH}(q_1,\ldots,q_D;t)
  =\sum_{n_1\ldots n_D}
  c_{n_1\ldots n_D}(t)
  \,\psi_{n_1}^{(1)}(q_1;t)\ldots\psi_{n_D}^{(D)}(q_D;t).
\end{equation}
From this expression, it is possible to obtain
the equations of motion for the one-dimen\-sional functions
$\psi_{n_i}^{(i)}(q_i;t)$ and for the correlation coefficients 
$c_{n_1\ldots n_D}(t)$.  
Without loss of generality we can choose the $\psi_{n_i}^{(i)}$'s
to be natural single-particle states.\cite{jan93}
They insure that we obtain the best approximation to the exact
$\Psi(q_1,\ldots,q_D;t)$ for a fixed number of configurations; i.e.,
they minimize the
expression $\langle\Delta\vert\Delta\rangle$ where 
\begin{equation}
  \Delta=\Psi-\Psi_{\rm MCTDH},
\end{equation}
and where $\Psi$ is the exact wave-function.
The natural single-particle functions are
eigenstates of the reduced density operators.
\begin{equation}
  \rho_j(q_j,q_j^\prime)
  =\int\!\!dq_1\ldots dq_{j-1}dq_{j+1}\ldots dq_D\\
  \quad\Psi(q_1\ldots q_{j-1}q_jq_{j+1}\ldots q_D)
  \Psi^*(q_1\ldots q_{j-1}q_j^\prime q_{j+1}\ldots q_D).
\end{equation}
The equations of motions for the natural single-particle states
can be obtained by differentiation of the eigenvalue equation:
\begin{equation}
  \int\!dq_j^\prime\rho_{i}(q_j,q_j^\prime)\psi_{n}^{(i)}(q_j^\prime)
  =\nu_{n}^{(i)}\psi_{n}^{(i)}(q_j).
\label {eq:rho}
\end{equation}
This gives us
\begin{equation}
  i\hbar{\partial\over\partial t}\psi_{n_j}^{(j)}
  =h_j\psi_{n_j}^{(j)}+\sum_{m_j}B_{n_jm_j}^{(j)}\psi_{m_j}^{(j)}
  +{\langle\tilde\psi_{n_j}^{(j)}\vert V\vert\Psi\rangle
   \over\langle\tilde\psi_{n_j}^{(j)}\vert
   \tilde\psi_{n_j}^{(j)}\rangle}
  -\sum_{m_j}{\langle\psi_{m_j}^{(j)}\tilde\psi_{n_j}^{(j)}
   \vert V\vert\Psi\rangle
   \over\langle\tilde\psi_{n_j}^{(j)}\vert
   \tilde\psi_{n_j}^{(j)}\rangle}
   \psi_{m_j}^{(j)},\label{eB}
\end{equation}
where
\begin{equation}
  B_{n_jm_j}^{(j)}={\langle\psi_{m_j}^{(j)}\tilde\psi_{n_j}^{(j)}
   \vert V\vert\Psi\rangle-\langle\Psi\vert V\vert
   \psi_{n_j}^{(j)}\tilde\psi_{m_j}^{(j)}\rangle\over
   \langle\tilde\psi_{n_j}^{(j)}\vert\tilde\psi_{n_j}^{(j)}\rangle
  -\langle\tilde\psi_{m_j}^{(j)}\vert\tilde\psi_{m_j}^{(j)}\rangle},
  \label{eE}
\end{equation}
and
\begin{equation}
  \tilde\psi_{n_j}^{(j)}=\sum_{n_1\ldots n_{j-1}}
  \sum_{n_{j+1}\ldots n_{D}}c_{n_1\ldots n_D}
  \psi_{n_1}^{(1)}\ldots\psi_{n_{j-1}}^{(j-1)}
  \psi_{n_{j+1}}^{(j+1)}\ldots\psi_{n_D}^{(D)},
\end{equation}
with the Hamiltonian $H$ given by
\begin{equation}
  H=\sum_{j=1}^Dh_j+V.
\end{equation}
The equations of motion for the coefficients
\begin{equation}
  c_{n_1\ldots n_D}=\langle\psi_{n_1}^{(1)}\ldots\psi_{n_D}^{(D)}
  \vert\Psi\rangle
\end{equation}
are again obtained by differentiation.
\begin{equation}
  i\hbar{d\over dt}c_{n_1\ldots n_D}
  =\langle\psi_{n_1}^{(1)}\ldots\psi_{n_D}^{(D)}\vert
    V\vert\Psi\rangle
  -\sum_{j=1}^Dc_{n_1\ldots n_{j-1}m_jn_{j+1}\ldots n_D}
  B_{m_jn_j}^{(j)}.
  \label{eC}
\end{equation}
Equations~(\ref{eB}) and (\ref{eC}) are a particular form
of the more general equations obtained from a time-dependent variational
principle.\cite{man92}
They conserve the norm of the wave-function and the mean energy
of a time-independent Hamiltonian.
The resulting system of first-order differential equations, 
has to be solved with a general-purpose integrator.
We used the variable-order variable-step Adams method, as implemented in
the NAG library.\cite{nag93} This method gave good convergence for all
described simulations. For the hardest simulation the method needed
195027 calculations of the time derivatives for 35000 atomic units
simulation time, which took around two days CPU-time on an 166 MHz. SUN
Ultra SPARC. 
The singularities in Eqs.\ (\ref{eB}), (\ref{eE}), and (\ref{eC}) have
been treated numerically by the regularization procedure described
in Ref.~\onlinecite{jan93}.

The natural single-particle states have some small advantages over other
possible choices of single-particle states.
The most important one is that one can directly see from 
$\langle\tilde\psi_{n}^{(i)}\vert\tilde\psi_{n}^{(i)}\rangle$ how well
$\Psi_{\rm MCTDH}$ approximates the exact wave-function.
How much a natural single-particle functions contributes to the wave-function
is given by the eigenvalue $\nu_n^{(i)}$
of the reduced density matrix.
In an approximate MCTDH simulation
$\langle\tilde\psi_{n}^{(i)}\vert\tilde\psi_{n}^{(i)}\rangle$
is an approximation for this exact eigenvalue.
The natural single-particle functions are also convenient for interpreting the
results of a simulations.
%%%%%%%%%%%%%%%%%%%%%%%%%%%%%%%%%
% The Potential Energy Surfaces %
%%%%%%%%%%%%%%%%%%%%%%%%%%%%%%%%%
\subsection{The Potential Energy Surfaces}
The PESs we used can all be written as
\begin{equation}
  \label{Vgen}
  V_{\rm total}=V_{\rm intra}+V_{\rm surf},
\end{equation}
where $V_{\rm intra}$ is the intramolecular PES and $V_{\rm surf}$ is the
interaction with the surface. For the $V_{\rm intra}$
we looked at four different types of PESs.

\subsubsection{A harmonic potential}

The first one is completely harmonic. We have used normal mode
coordinates for the internal vibrations, because these are coupled only
very weakly.
In the harmonic approximation this coupling is even absent so that we
can write $V_{\rm intra}$ as 
\begin{equation}
  \label{Vharm}
  V_{\rm intra}=V_{\rm harm}= {1\over 2}\sum_{i=2}^{10}k_iX_i^2,
\end{equation}
the summation is over the internal vibrations, $X_i$'s are mass-weighted
displacement coordinates and $k_i$ are mass-weighted force constants. (see
Table \ref{tab:gen} for definitions and values); ($X_1$ is the
mass-weighted overall translation along the surface normal).\cite{wil55}
The force constants have been obtained by fitting them on the
experimental vibrational frequencies of CH${}_4$ and
CD${}_4$.\cite{gray79,lee95} 

We have assumed that the interaction with the surface is only through
the hydrogen atoms that point towards the surface. We take the $z$-axis
as the surface normal. In this case the surface PES is given by
\begin{equation}
  \label{gen_Vsurf}
  V_{\rm surf}={A \over{N_H}}\sum_{i=1}^{N_H} e^{-\alpha z_i},
\end{equation}
where $N_H$ is the number of hydrogens that points towards the surface,
$\alpha$=1.0726 atomic units and $A$=6.4127 Hartree. These parameters
are chosen to give the same repulsion as the PES that has been used in
an MCTDH wavepacket simulation of CH${}_4$ dissociation.\cite{jan95} 

\begin{figure}[H]
  \epsfig{file=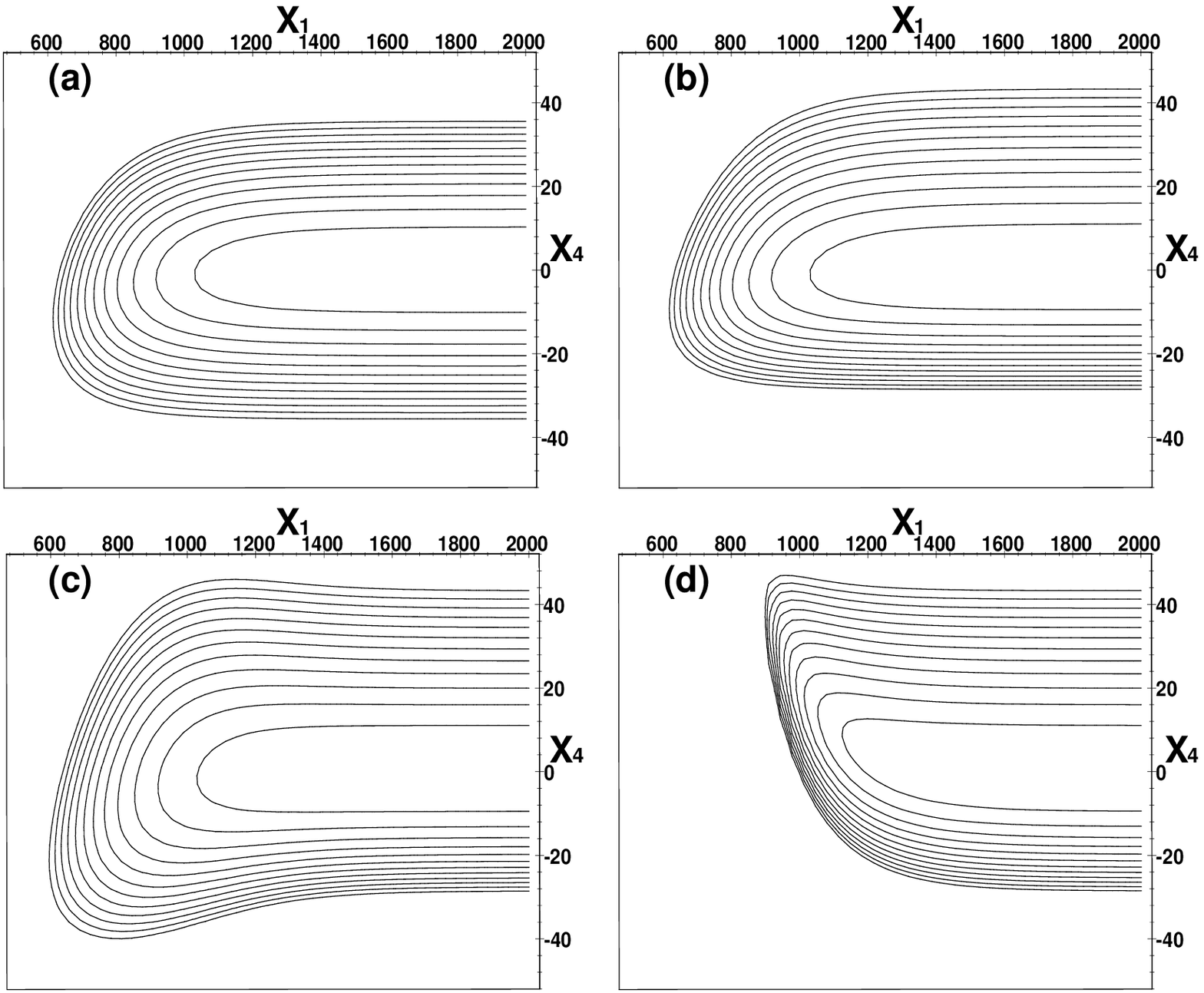, height=10cm}
    \caption{Contour plots in the energy range from 0 to 0.12 Hartree of
      a cross-section in the orientation with one hydrogen atom pointing
      towards the surface for the translational mode $X_1$ and the
      $\nu_3$ asymmetrical mode $X_4$ (an $a_1$ mode in C${}_{3v}$),
      both in atomic displacement, which show well the different
      behaviours of the four PESs; (a) harmonic
      PES, (b) an anharmonic intramolecular PES, (c)
      intramolecular PES with weakening C--H bond, and (d)
      intramolecular PES with elongation of the C--H bond. All
      other coordinates set to zero.} 
    \label{fig:potdif}
\end{figure}

If we write $V_{\rm surf}$ in terms of normal mode coordinates, then we
obtain for one hydrogen pointing towards the surface 
\begin{equation}
  \label{vsurfone}
  V_{\rm surf}= A
  e^{-\alpha_1X_1}e^{-\alpha_2X_2}e^{-\alpha_3X_3}e^{-\alpha_4X_4},
\end{equation}
where $A$ as above, and $\alpha$'s as given in Table \ref{tab:alpha}.
$X_2$, $X_3$ and $X_4$ correspond all to $a_1$ modes of the C${}_{3v}$
symmetry. There is no coupling between the modes $X_5$ to $X_{10}$ in the
$V_{\rm surf}$ part of the PES, which are all $e$ modes of the C${}_{3v}$
symmetry. Fig.~\ref{fig:potdif}(a) shows a contour plot of the
cross-section of the total harmonic PES with one hydrogen
pointing towards the surface in the translational mode $X_1$ and the
$\nu_3$ asymmetrical stretch mode $X_4$.

For two hydrogens we obtain
\begin{eqnarray}
  \label{vsurftwo}
  V_{\rm surf}=A&&e^{-\alpha_1X_1}e^{-\alpha_2X_2}
  e^{-\alpha_3X_3}e^{-\alpha_4X_4}e^{-\alpha_5X_5}\\
  \times{1\over2}\Big[
  &&e^{\beta_{3}X_7}
    e^{-\beta_{3}X_8}e^{-\beta_{5}X_9}e^{\beta_{5}X_{10}}
    \nonumber\\
  +&&e^{-\beta_{3}X_7}
    e^{\beta_{3}X_8}e^{\beta_{5}X_9}e^{-\beta_{5}X_{10}}
    \Big],\nonumber
\end{eqnarray}
with $A$ again as above, $\alpha$'s and $\beta$'s as given in Table
\ref{tab:alpha}.
$X_2$, $X_3$, $X_4$ and $X_5$ correspond all to $a_1$ modes of C${}_{2v}$. 
$X_7$, $X_8$, $X_9$ and $X_{10}$ correspond to $b_1$ and $b_2$ modes of
C${}_{2v}$. $X_6$ corresponds to the $a_2$ mode of C${}_{2v}$ and
has no coupling with the other modes in $V_{\rm surf}$.

For three hydrogens we obtain
\begin{eqnarray}
  \label{vsurfthree}
  V_{\rm surf}=A&&e^{-\alpha_1X_1}e^{-\alpha_2X_2}
  e^{-\alpha_3X_3}e^{-\alpha_4X_4}\\
  \times{1\over 3}\Big[
  &&e^{\beta_{1}X_5}e^{\beta_{2}X_6}e^{-\beta_{3}X_7}
    e^{-\beta_{4}X_8}e^{\beta_{5}X_9}e^{\beta_{6}X_{10}}
    \nonumber\\
  +&&e^{\beta_{1}X_5}e^{-\beta_{2}X_6}e^{-\beta_{3}X_7}
    e^{\beta_{4}X_8}e^{\beta_{5}X_9}e^{-\beta_{6}X_{10}}
    \nonumber\\
  +&&e^{-2\beta_{1}X_5}e^{2\beta_{3}X_7}e^{-2\beta_{5}X_9}
    \Big],\nonumber
\end{eqnarray}
$X_2$, $X_3$ and $X_4$ corresponds to $a_1$ modes in the C${}_{3v}$ symmetry.
Because these last six coordinates correspond to degenerate $e$ modes of
the C${}_{3v}$ symmetry, the $\beta$ parameters are not unique.

%
% Morse Potential
%
\subsubsection{An anharmonic intramolecular potential}
\label{sec:morse}

Even though we do not try to describe the dissociation of methane in
this paper, we do want to determine which internal vibration might be
important for this dissociation. The PES should
at least allow the molecule to partially distort as when
dissociating. The harmonic PES does not do this. A number of
changes have therefor been made. The first is that we have describe the
C--H bond by a Morse PES.
\begin{equation}
  \label{Vmorse_n}
  V_{\rm Morse}=D_e \sum_{i=1}^4\Big[1-e^{-\gamma \Delta r_i}\Big]^2 ,
\end{equation}
where $D_e=0.1828$ Hartree (the dissociation energy of methane in the
gas-phase) and $\Delta r_i$ the change in bond length from the
equilibrium distance. $\gamma$ was
calculated by equating the second derivatives along one bond of the
harmonic and the Morse PES. 
If we transform Eq.\ (\ref{Vmorse_n}) back into normal mode
coordinates, we obtain
\begin{equation}
  \label{Vmorse_d}
  V_{\rm Morse}=D_e
  \sum_{i=1}^4\Big[1-e^{\gamma_{i2}X_2}e^{\gamma_{i3}X_3}e^{\gamma_{i4}X_4}
  e^{\gamma_{i7}X_7}e^{\gamma_{i8}X_8}e^{\gamma_{i9}X_9}
  e^{\gamma_{i,10}X_{10}}\Big]^2,
\end{equation}
with $D_e$ as above. $\gamma$'s are given in Tables \ref{tab:gammaone}
and \ref{tab:gammatwo}. Note that, although we have only changed the
PES of the bond lengths, the $\nu_4$ umbrella modes are also
affected. This is because these modes are not only bending, but also
contain some changes of bond length.

The new intramolecular PES now becomes
\begin{equation}
  \label{Vintra_morse}
  V_{\rm intra}=V_{\rm harm}+V_{\rm Morse}-V_{\rm corr}, 
\end{equation}
where $V_{\rm harm}$ is as given in Eq.\  (\ref{Vharm}) and $V_{\rm corr}$ is
the quadratic part of $V_{\rm Morse}$, which is already in $V_{\rm harm}$. 
Fig.~\ref{fig:potdif}(b) shows a contour plot of the
cross-section of this total anharmonic PES with one hydrogen
pointing towards the surface in the translational mode $X_1$ and the
$\nu_3$ asymmetrical stretch mode $X_4$.
(We would like to point out that there are, of course, various
anharmonic PESs for methane in the 
literature. There are two reasons why we haven't use them. First, these
PESs are not in appropriate form to use then with the MCTDH
method.\cite{man92,jan93} Second, these PESs are generally quite
complicated. We prefer to keep it as simple as possible, because at this
moment we're only interested in qualitative effects.)

%
% Weak Potential
%
\subsubsection{Intramolecular potential with weakening C--H bonds}
\label{sec:weak}

When the methane molecule approach the surface the overlap of substrate
orbitals and anti-bonding orbitals of the molecule weakens the C--H bonds.
We want to include this effect for the C--H bonds of the hydrogens
pointing towards the surface. We have redefined the $V_{\rm Morse}$ 
given in Eq.\ (\ref{Vmorse_d}) and also replace it in
Eq.\ (\ref{Vintra_morse}). A sigmoidal function is used to switch from the
gas phase C--H bond to a bond close to the surface. We have used the
following, somewhat arbitrary, approximations. 
(i) The point of inflection should be at a reasonable distance from the
surface. It is set to the turnaround point for a rigid methane molecule
with translation energy 93.2 kJ/mol plus twice the fall-off distance of
the interaction with the surface. (ii) The depth of the PES of the C--H
bond is 480 kJ/mol in the gas phase, but only 93.2 kJ/mol near the
surface. The value 93.2 kJ/mol corresponds to the height of the
activation barrier used in our dissociation.\cite{jan95} (iii) The
exponential factor is the same as for the interaction with the surface. 

If we transform to normal-mode coordinates for the particular
orientations, we then obtain
\begin{equation}
  \label{Vweak_d}
  V_{\rm weak}=D_e
  \sum_{i=1}^4W_i\Big[1-e^{\gamma_{i2}X_2}e^{\gamma_{i3}X_3}
  e^{\gamma_{i4}X_4}
  e^{\gamma_{i7}X_7}e^{\gamma_{i8}X_8}e^{\gamma_{i9}X_9}
  e^{\gamma_{i,10}X_{10}}\Big]^2,
\end{equation}
where $W_i=1$ for non-interacting bonds and
\begin{equation}
  \label{Wi}
  W_i= { {1 + \Omega e^{-\alpha_1X_1 + \omega}} \over {1 + 
  e^{-\alpha_1X_1 + \omega}} }  
\end{equation}
for the interacting bonds pointing towards the surface.
$\alpha_1$ is as given in Table \ref{tab:alpha}, $\gamma$'s are given in
Tables \ref{tab:gammaone} and \ref{tab:gammatwo}, $\Omega=1.942
\cdot10^{-1}$ and $\omega=7.197$.
Fig.~\ref{fig:potdif}(c) shows a contour plot of the cross-section of this
total anharmonic PES with weakening C--H bond for the orientation
with one hydrogen pointing towards the surface in the translational mode
$X_1$ and the $\nu_3$ asymmetrical stretch mode $X_4$.

%
% The Shift PES
%
\subsubsection{Intramolecular potential with elongation of the C--H bond}
\label{sec:shift}

A weakened bond generally has not only a reduced bond strength, but also
an increased bond length.
We include the effect of the elongation of the C--H
bond length of the hydrogens that point towards the surface due to
interactions with the surface.
We have redefined the $V_{\rm Morse}$ given in Eq.\ (\ref{Vmorse_d}) and
also replace it in Eq.\ (\ref{Vintra_morse}) for this type of PES.
We have used the following approximations:
(i) The transition state, as determined by Refs.\onlinecite{bur93b,bur93c}, has
a C--H bond that is 0.54 {\AA} longer than normal. This elongation
should occur at the turn around point for a rigid methane molecule with
a translation energy of 93.2 kJ/mol. 
(ii) The exponential factor is again the same as for the interaction with the
surface. 

\begin{figure}[H]
  \epsfig{file=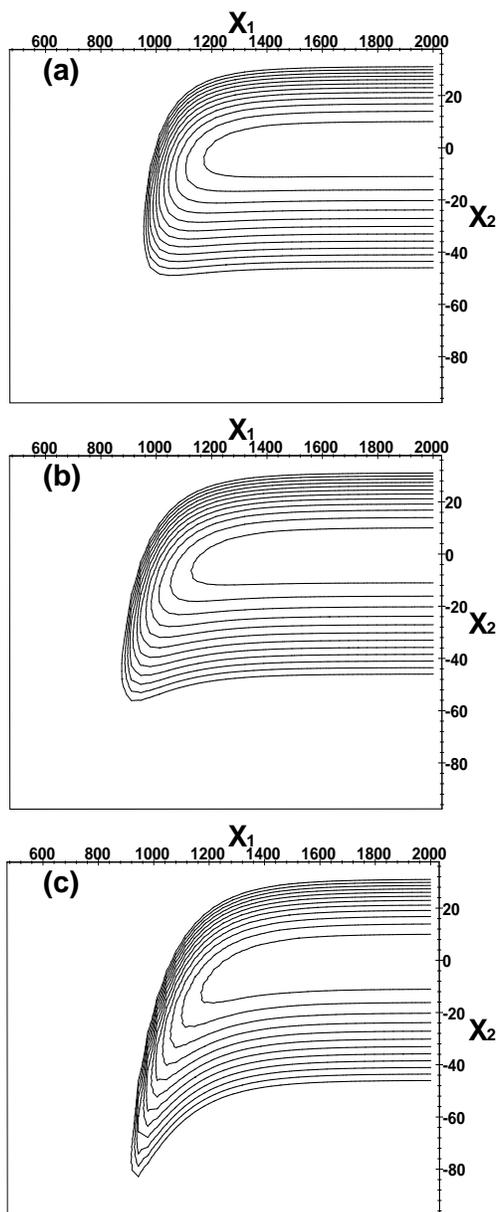, height=16cm}
    \caption{Contour plots in the energy range from 0 to 0.12 Hartree
      of a cross-section of the PES with  
      elongation of the C--H bond in the translational $X_1$
      and the $\nu_1$ symmetrical stretch mode $X_2$, both in
      atomic displacement; 
      (a) one, (b) two, and (c) three hydrogens pointing towards the
      surface. All other coordinates set to zero.} 
    \label{fig:symshift}
\end{figure}

If we transform to normal-mode coordinates for the particular
orientations, then we obtain
\begin{equation}
  \label{Vshift_d}
  V_{\rm shift}=D_e
  \sum_{i=1}^4\Big[1-e^{\gamma_{i2}X_2}e^{\gamma_{i3}X_3}e^{\gamma_{i4}X_4}
  e^{\gamma_{i7}X_7}e^{\gamma_{i8}X_8}e^{\gamma_{i9}X_9}
  e^{\gamma_{i,10}X_{10}}\exp[S_ie^{-\alpha_1X_1}]\Big]^2 ,
\end{equation}
where $\alpha_1$ is as given in Table \ref{tab:alpha}, $\gamma$'s are
given in Tables \ref{tab:gammaone} and \ref{tab:gammatwo}. For
orientation with one hydrogen towards the 
surface we obtain; $S_1=2.942 \cdot 10^{2}$ and $S_2=S_3=S_4=0$,
with two hydrogens; $S_1=S_2=0$ and $S_3=S_4=1.698 \cdot 10^{2}$,
and with three hydrogens; $S_1=0$ and $S_2=S_3=S_4=2.942 \cdot
10^{2}$. 
Fig.~\ref{fig:potdif}(d) shows a contour plot of the cross-section of this
total anharmonic PES with elongation of the C--H bond for the
orientation with one hydrogen pointing towards the surface in the
translational mode $X_1$ and the $\nu_3$ asymmetrical stretch mode
$X_4$, and Fig.~\ref{fig:symshift} shows a contour plot of a
cross-section in the translational $X_1$  and the $\nu_1$ symmetrical
stretch mode $X_2$ in the three different orientations.

Finally we like to note that the monotonic behaviour of all PES types in
the translational mode does not contradict with the fact that the
dissociative adsorption of methane is activated. The activation barrier
for dissociative adsorption in the translational mode is situated in a
region with very high excitations of the stretch modes, which we do not
reach in these simulations.

\subsection{Initial States}
\label{sec:states}

All initial states in the simulations start with the vibrational
ground state. The initial translational part $\psi^{({\rm tr})}$ is
represented by a Gaussian wave-packet, 
\begin{equation}
  \label{Gausswave}
  \psi^{({\rm tr})}(X_1)=(2\pi\sigma^2)^{-1/4}
  \exp\left[-{(X_1-X_0)^2\over 4\sigma^2}+
  iP_1X_1\right],
\end{equation}
where $\sigma$ is the width of the wave-packet (we used $\sigma=320.248$
atomic units), $X_0$ is the initial position (we used $X_0=11\sigma$,
which is far enough from the surface to observe no repulsion) and $P_1$
is the initial momentum. We performed simulations in the energy range of
32 to 96 kJ/mol. We here present only the results of 96 kJ/mol
(equivalent to $P_1=0.2704$ atomic units), because they showed the most
obvious excitation probabilities for $V_{\rm Morse}$. We used seven
natural single-particle states, 512 grid points and a grid-length of
$15\sigma$ for the translational coordinate. With this grid-width we can
perform simulation with a translational energy up to 144 kJ/mol.

Gauss-Hermite discrete-variable representations (DVR) \cite{light85}
were used to 
represent the wavepackets of the vibrational modes.
We used for all simulations 5 DVR points for the $\nu_2$ bending modes and 8
DVR points for the $\nu_4$ umbrella, $\nu_3$ asymmetrical
stretch, and $\nu_1$ symmetrical stretch mode for an numerical exact
integration, except for the simulations with $V_{\rm shift}$,
where we used 16 DVR points for the $\nu_1$ symmetrical stretch mode,
because of the change in the equilibrium position.

We did the simulation with one hydrogen pointing towards the
surface in eight dimensions, because the $\nu_2$ bending modes $X_5$ and
$X_6$ do not couple with the other modes. We needed four natural
single-particle states for modes $X_2$, $X_3$ and $X_4$, and just one
for the others. So the number of configurations was $7^1 \cdot 4^3 \cdot
1^4 = 448$.  
The simulation with two hydrogens pointing towards the surface was
performed in nine dimensions. One of the $\nu_2$ bending mode ($X_6$)
does not couple with the other modes, but for the other mode $X_5$ we
needed four natural single-particle states. The number of configurations was
$7^1 \cdot 4^4 \cdot 1^4 = 1792$, because we needed the same number of
natural single-particle states as mentioned above for the other modes.
We needed ten dimensions to perform the simulation with three hydrogens
pointing towards the surface. We used here one natural single-particle
state for the modes $X_5$ to $X_{10}$ and four natural single-particle
states for $X_2$ to $X_4$, which gave us $7^1 \cdot 4^3 \cdot 1^6 = 448$
configurations.

%%%%%%%%%%%%%%%%%%%%%%%%%%
% RESULTS AND DISCUSSION %
%%%%%%%%%%%%%%%%%%%%%%%%%%
\section{Results and Discussion}

We found that the scattering is predominantly elastic. The elastic
scattering probability is larger than 0.99 for all orientations and PESs
at a translational energy of 96 kJ/mol, except for the PES with $V_{\rm
shift}$ and three hydrogens pointing towards the surface for which it is 0.956.
This agrees with the observation that the translation-vibration coupling
is generally small.\cite{lev87}

If we would have wanted to determine the role of the internal vibrations
from the scattering probabilities, we would have to do quite accurate
simulations. We have opted instead to look at the molecule when it hits
the surface, which enables us to obtain good results with much less
costly simulations.

\subsection{Excitation Probabilities}

The surface PES has C${}_{3v}$ (with one or three hydrogens
towards the surface) or C${}_{2v}$ symmetry (with two hydrogens towards
the surface). If we expand this PES in a Taylor-series of internal
vibrations, we see that the linear terms contain only those vibrations
that transform as $a_1$ in C${}_{3v}$, respectively, C${}_{2v}$. These
are therefore easier to excite than others; We did not find any
appreciable excitation of $e$ modes of C${}_{3v}$ and the $a_2$, $b_1$,
and $b_2$ modes of C${}_{2v}$.

We will not present results of the simulations with the harmonic PES,
because they give almost the same excitation probabilities as the PES
with $V_{\rm Morse}$. 
The maximum excitation probabilities at the surface for the PES with
$V_{\rm Morse}$ are given in Table \ref{tab:exprop}. We have observed
the highest excitations for this PES in the $\nu_4$ umbrella
and $\nu_2$ bending modes in the orientation with two hydrogens, and
for the $\nu_4$ umbrella mode in the orientation with three
hydrogens pointing towards the surface. 
The excitation probabilities for the $\nu_1$ and $\nu_3$
stretch modes at this orientations are a factor of magnitude lower.  

We have observed higher excitation probabilities for the $\nu_1$
and $\nu_3$ stretch modes in the one hydrogen orientation. The
excitation probability of the $\nu_4$ umbrella
mode is here lower then for the other orientations, but still in the
same order as the stretch modes for this orientation.
This can be explained by the values of the $\alpha$'s of $V_{\rm surf}$ (see
Table \ref{tab:alpha}) and the force constants of $V_{\rm harm}$ (see
Table \ref{tab:gen}), because $V_{\rm harm}$ is approximately $V_{\rm Morse}$
for the ground state. The force constants of $V_{\rm intra}$ for the
$\nu_1$ symmetrical stretch stretch and $\nu_3$ asymmetrical
stretch are of the same order, but the $\alpha$ parameter in $V_{\rm
  surf}$ in the orientation with one hydrogen pointing towards the
surface for the $\nu_3$ asymmetrical stretch is around twice as large
as for the $\nu_1$ symmetrical stretch, which explains why
$\nu_3$ is more excited as $\nu_1$. The surface repulsion on the
$\nu_4$ umbrella mode is even three times lower then on the
$\nu_1$ symmetrical stretch mode, but the force constant is also
much lower. It results in a little more excitation of the $\nu_4$
umbrella then in the $\nu_1$ symmetrical stretch. 

For the orientation with two hydrogens pointing towards the surface the
repulsion on the vibrational modes is for all modes in the same order
($\nu_2$ bending a little higher), so the difference in excitation
probabilities correlate here primarily with the force constants. The force
constants of the $\nu_2$ bending and $\nu_4$ umbrella modes are
of the same order, as are those of $\nu_1$ and $\nu_3$. The stretch
force constants are higher, however, so that the excitation
probabilities are lower. 
The repulsion on the $\nu_4$ umbrella mode is the largest in the
orientation with three hydrogens pointing towards the surface. The force
constants is lower for this modes then for the $\nu_1$ and
$\nu_3$ stretch modes, so the primary excitation is seen in the
$\nu_4$ umbrella mode. 

For the orientation with three hydrogens pointing towards the
surface the repulsion on the $\nu_4$ umbrella mode is the highest in
combination with a low force constant, so this mode has a much higher
excitation probability than the $\nu_1$ and $\nu_3$ stretch modes.
Another interesting detail is that the
$\alpha_3$ parameter is higher for the orientation with two than 
with three hydrogens pointing toward the surface, but the
excitation probabilities for the $\nu_4$ umbrella in this
orientations are equal. This is caused by a coupling between
the excitation of the $\nu_2$ bending and the $\nu_4$ umbrella
mode in the orientation with two hydrogens pointing towards the surface.

We observed with $V_{\rm weak}$ (see Table \ref{tab:exprop}) that all
excitation probabilities become much higher than with $V_{\rm Morse}$,
except for the $\nu_4$ umbrella mode with two hydrogens towards the
surface, which stays almost the same.  
It is caused by the fact that, although the $\nu_1$ symmetrical stretch
and $\nu_3$ asymmetrical stretch contribute almost completely to
$V_{\rm Morse}$, the $\nu_4$ umbrella does so just for a small part. The
$\nu_4$ umbrella contributes primarily, and the $\nu_2$ bending
completely, to the harmonic terms of intramolecular PES
$V_{\rm intra}$. $V_{\rm weak}$ gives only a lowering in the $V_{\rm
  Morse}$ terms of $V_{\rm intra}$, so we should expect primarily an
higher excitation probability in the $\nu_1$ and $\nu_3$ stretch
modes. This will also cause a higher excitation probability of
the other modes, because the turn-around point will be some what later,
which give effectively more repulsion on the other modes.

We also observed that for both stretches the excitation
probabilities shows the following trend; three hydrogens $<$ two hydrogens
$<$ one hydrogen pointing towards the surface. For one hydrogen pointing
towards the surface the excitation probability of the $\nu_3$
asymmetrical stretch is around twice that of the $\nu_1$
symmetrical stretch, they are almost equal for two hydrogens, and for
three hydrogens the $\nu_1$ is twice the $\nu_3$
stretch. Some of these trends can also be found for the PES with $V_{\rm
  Morse}$, but are not always that obvious. They follow the same trends
as the $\alpha$ parameters of $V_{\rm surf}$ in Table \ref{tab:alpha}.

$V_{\rm shift}$ (see Table \ref{tab:exprop}) gives also higher
excitation probabilities for almost all modes than $V_{\rm Morse}$, 
but for the $\nu_2$ bending mode in the two hydrogens pointing
towards the surface orientation it became lower. This means that the
repulsion of the surface is here caused for a large part by the $V_{\rm
  shift}$ terms (see Eq.\  (\ref{Vshift_d})), where the
$\nu_2$ bending modes don't contribute. It is also in agreement with
the fact that we have observed the excitation maximum earlier, so the
$V_{\rm surf}$  repulsion on the $\nu_2$ bending modes will be
lower. Fig.~\ref{fig:potdif} shows that the repulsion for the PES with
$V_{\rm shift}$ is stronger than for the other three PESs.

The excitations in the $\nu_1$ and $\nu_3$ stretch modes are
extremely high with $V_{\rm shift}$, because they contribute strongly to
the C--H bond elongation.  
The trend in the excitation probabilities of the $\nu_1$ symmetrical
stretch mode is caused by the number of bonds pointing towards the surface,
because the $\gamma_{i2}$ parameters are for all bonds and orientations
the same, so only the strength of the coupling with the shift-factor
dominates. This effect is illustrated in Fig.~\ref{fig:symshift}.
We didn't observe this trend in the $\nu_3$ asymmetrical
stretch mode, because there is also difference between the $\gamma_{i4}$
parameters as can be seen in Table \ref{tab:gammaone} and
\ref{tab:gammatwo}.

\subsection{Structure Deformation}
\label{sec:strdef}

\begin{figure}[H]
  \epsfig{file=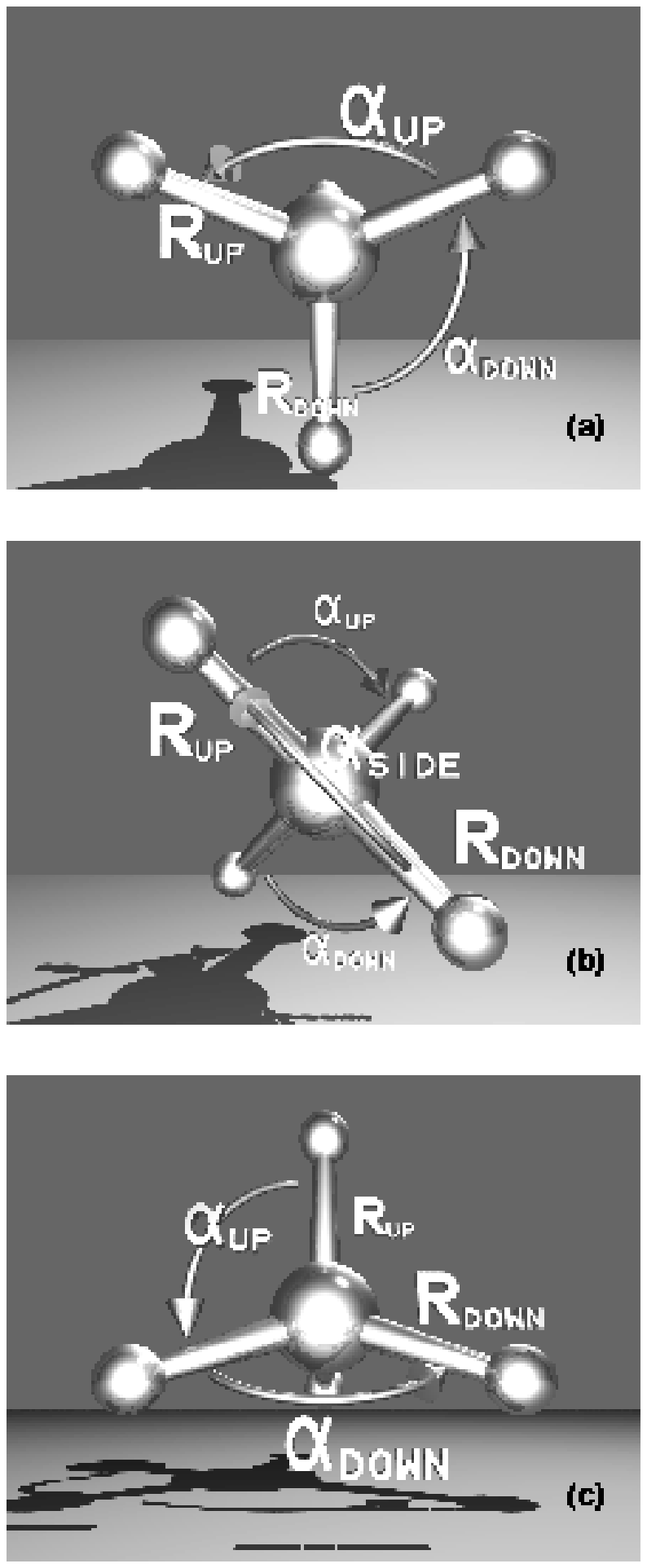, height=17cm}
    \caption{Schematic representation of the $R_{\rm down}$ and
      $R_{\rm up}$ 
      bonds, and the $\alpha_{\rm down}$, $\alpha_{\rm side}$ and
      $\alpha_{\rm up}$ angles for the three orientations: (a) one, (b)
      two, and (c) three hydrogens pointing towards the surface.}
    \label{fig:orient}
\end{figure}

If we put the methane molecule far from the surface, then all PESs are
identical. For this situation we calculated a bond distance of 1.165
{\AA} and bond angle of 109.5 degrees in the ground state. 
The results of the maximum structure deformations are shown in the Table
\ref{tab:strdef}. Fig. \ref{fig:orient} shows the names of the bonds and
angles for the three orientations.

We observed that the PESs with $V_{\rm Morse}$ and $V_{\rm weak}$ give
again the same trends, but that for the PES with $V_{\rm weak}$ these
trends are much stronger. This is in agreement with the observations
discussed for the excitation probabilities above. The deformations for
the PES with $V_{\rm shift}$ are dominated
by the change of the bond distances for the bonds which are pointing
towards the surface. These bonds become longer for all orientations in
the same order. So there is no orientational effect and this will be
probably caused completely by the  $V_{\rm shift}$ terms (see
Eq.\  (\ref{Vshift_d})). 

For the orientation with one hydrogen pointing towards the surface (see
Table \ref{tab:strdef}), we observed that the bond pointing towards the
surface becomes shorter, as expected, for the PES with $V_{\rm Morse}$
and even more with $V_{\rm weak}$. This is caused by the repulsion of
the $V_{\rm surf}$ terms, which works in the direction of the bond
axes. We observed also some small bond angle deformation. This is
probably a secondary effect of the change in the bond distance and
correlates to excitation of the $\nu_4$ umbrella mode.
Remarkably, for $V_{\rm shift}$ the bond length increases. Clearly the
effect of the change in equilibrium bond length is more effective than
the repulsion with the surface. The shift effect may be somewhat too
large, but the observed change (0.232 {\AA}) is much lower
than the shift of the equilibrium (0.54 {\AA}) on which we fitted
$V_{\rm shift}$. 

We also observed for all orientations with $V_{\rm shift}$ an earlier
turn-around point, which is in agreement with the earlier observed
excitation maximum. This is caused by the extra repulsion contribution
in the $V_{\rm shift}$ terms and by the longer bond length; a longer
bond gives a higher repulsion for the same position of the center of
mass. Fig. \ref{fig:potdif} shows this effect in the $\nu_3$
asymmetrical stretch mode.

The orientation with two hydrogens pointing towards the surface (see
Table \ref{tab:strdef}) gives also shorter bond distances for the
bonds pointing towards the surface for the PESs with $V_{\rm Morse}$ and
$V_{\rm weak}$, although very little in comparison to the one hydrogen
orientation. We expected this, because just half of the repulsion is now
in the direction of the bonds. 
The other half is perpendicular on the C--H bonds and this makes the
bond angle larger. The bond angles between the bonds pointing towards
the surface show a quite large deformation. This was already expected
from the excitation of the $\nu_2$ bending and $\nu_4$ umbrella
modes. 

We also observed that a smaller bond length of the bonds pointing
towards the surface correlates with a larger bond angle of the angles
between the bonds pointing towards the surface and a higher excitation
probability of the $\nu_2$ bending mode. This can be explained as
follows. When the bonds become shorter the center of mass can come
closer to the surface. Because of this there will be more repulsion on
the $\nu_2$ bending mode, which will cause a larger bond angle
deformation.

Most of the deformations of the
other bond angles can be explained as an indirect effect of the
deformation of this bond angle, but the deformation of the angle between
the bonds pointing away from the surface which was seen at the PES with
$V_{\rm shift}$ must be caused by the excitation of the $\nu_4$ umbrella
mode. 

We observed almost no change in bond length for the PESs with $V_{\rm
Morse}$ and $V_{\rm weak}$ at the three hydrogens towards the surface
orientation (see Table \ref{tab:strdef}). Almost all energy is absorbed
by the $\nu_4$ umbrella mode, so this gives quite large bond angle
deformations. The bond angle deformation is larger for the PES with
$V_{\rm weak}$ than with $V_{\rm Morse}$ PES; in agreement with higher
excitation probability for the PES with $V_{\rm weak}$. The PES with
$V_{\rm shift}$  has around the same excitation probabilities for the
$\nu_4$ umbrella  mode as the PES with $V_{\rm weak}$, but the bond
angle deformation is the same as for the PES with $V_{\rm Morse}$. This
may be caused by the higher excitations of the $\nu_1$ and $\nu_3$
stretch modes and the longer bonds pointing towards the surface at the
PES with $V_{\rm shift}$, which can make it harder to deform the bond
angles.

\subsection{Dissociation models}
\label{sec:diss}

Several dissociation mechanisms for direct methane dissociation on
transition metals have been suggested. All proposals can be related to
two main ideas. One of them is the breaking of a single C--H bond in the
initial collision.\cite{lee87,lun91} This model is most suggested in
literature and also all wavepacket simulations have focussed on the effects
of this model.\cite{jan95,lun91,lun92,lun95,car98} The other mechanism
is often called ``splats'' and suggests that the critical requirement
for methane dissociation is angular deformation of methane which allows
a Ni--C bond to form.\cite{lee87} Even though we did not try to describe
the dissociation itself, we would like to discuss the implication of our
simulation for the dissociation.

The ``splats'' model seems easiest to discuss. Angular deformation is
related to the $\nu_2$ bending and the $\nu_4$ umbrella mode. The
excitation probabilities in Table~\ref{tab:exprop} seem to indicate that
these modes are easy to excite, and that angular deformation should be
large. These excitation probabilities are misleading, however.
Table~\ref{tab:strdef} shows that, although the excitation probabilities
depend on the PES, the changes in the bond angles
do not. Moreover, these changes are quite small. They are largest for
the orientation with two hydrogen atoms pointing towards to surface, for
which they are about $8\,$degrees at most. This seems much too small to
enable the formation of a Ni--C bond.  A previous estimate of the bond
angle deformation gave an energy of $68.1\,$kJ/mol to get three
hydrogens and the carbon in one plane. This correspond to a bond angle
of $120\,$degrees.\cite{lee87} We find with three hydrogens pointing
towards the surface that the bond angle only changes to about
$112\,$degrees at the higher energy of $96\,$kJ/mol. The reason for this
difference is that for the older estimate it was assumed that all
translational energy can be used to deform the bond angle, whereas
Tables~\ref{tab:exprop} and \ref{tab:strdef} clearly show that this is
not correct.

The excitation probabilities for the stretch modes depend strongly on
the orientation of the molecule and on the PES. For
the PESs without a change of the equilibrium bond
length the excitation probabilities are only appreciably with one
hydrogen pointing towards the surface. This is, however, not a
favourable orientation for the dissociation. Moreover,
Table~\ref{tab:strdef} shows that in those cases the repulsion with the
surface shortens the C--H bond. The same holds for the orientation with
two hydrogens pointing towards the surface, although to a lesser extent,
whereas the bond length changes hardly at all with three hydrogens
pointing towards the surface. This indicates that the intramolecular
PES needs the bond elongation to overcome the
repulsion of the surface that shortens the C--H bond, and to get
dissociation. This agrees completely with electronic structure
calculations that yield a late barrier for dissociation with a very
elongated C--H bond.\cite{bur93a,kra96} The large excitation
probabilities for the PES with the elongated
equilibrium bond length should not be over-interpreted, however. They are
to be expected even if the molecule stays in its (shifted) vibrational
ground state. More telling is that for this PES we do find at
least some inelastic scattering.

%
%%%%%%%%%%%%%%%
% CONCLUSIONS %
%%%%%%%%%%%%%%%
\section{Conclusions}

We have done wavepacket simulations on the scattering of methane from a
flat Ni(111) model surface with a fixed orientation with one, two, or
three hydrogens pointing towards the surface. We used the MCTDH method
and four different model PESs for each orientation.
We used a translational energy of up to $96\,$kJ/mol and all internal
vibrations in the ground state. The scattering was in all cases
predominantly elastic.

When the molecule hits the surface, we always observe vibrational
excitations of the $\nu_4$ umbrella and $\nu_2$ bending modes,
especially in the orientations with two or three hydrogens pointing
towards the surface. This is due to a favorable coupling that originates
from the repulsive interaction with the surface, and the low excitation
energies. Deformations of the molecule are predominantly in the bond
angles. The changes in the bond angles are, however, too small to allow
for the formation of a Ni--C bond, as suggested in the ``splats'' model
of methane dissociation.

Appreciable excitations of the $\nu_1$ and $\nu_3$ stretch modes when
methane hits the surface are only observed when one hydrogen atom points
towards the surface, or when the intramolecular PES has an
elongated equilibrium C--H bond length close to the surface. The
repulsion of the surface shortens the C--H bond. This can only be
overcome when the intramolecular PES incorporates the effect of a
longer equilibrium C--H bond length caused by overlap of occupied
surface orbitals with the antibonding orbitals of methane. This agrees
with quantum chemical calculations, which show a late barrier for
dissociation.

The simulations with these model PESs show that the internal
vibrations play an important role in the dissociation mechanism.
Excitation probabilities when the molecule hits the surface show how the
translational energy is converted into vibrational energy and it is
distributed over the internal modes. These probabilities vary strongly
with the PES. As only few internal vibrations contribute to the
dissociation, it is important to obtain more information on the real
PES for this system.

%
%
%%%%%%%%%%%%%%%%%%%%
% Acknowledgements %
%%%%%%%%%%%%%%%%%%%%
\section*{Acknowledgements}
We would like to thank SON (Stichting Scheikundig Onderzoek in
Nederland) and NWO (Nederlandse Organisatie voor
Wetenschappelijk Onderzoek) for their financial support. 

This work has been performed under the auspices of NIOK, the Netherlands
Institute for Catalysis Research, Lab Report No. TUE-98-5-02. 

{\it Copyright 1998 American Institute of Physics. This article may be
  downloaded for personal use only. Any other use requires prior
  permission of the author and the American Institute of Physics. This
  article appeared in J. Chem. Phys. [{\bf 109}, 1966 (1998)] and may
  be found at {\tt http://link.aip.org/link/?jcp/109/1966 }}.
\newpage

\begin{table}[H]
  \caption{Overview of the relations between the mass-weighted coordinates
  $X_i$; the force constants $k_i$ (in atomic units), the designation,
  and the symmetry in T${}_d$, C${}_{3v}$ and C${}_{2v}$.} 
  \label{tab:gen}
  \begin{tabular}{r l l c c c}
    \hline
    $i$ & $k_i$ & designation & T${}_d$ & C${}_{3v}$ & C${}_{2v}$ \\
    \hline
    $1$ &                      & translation & $t_2$ & $a_1$ & $a_1$ \\ 
    $2$ & $1.780\cdot 10^{-4}$ & $\nu_1$; symmetrical stretch & $a_1$ &
    $a_1$  & $a_1$ \\ 
    $3$ & $3.599\cdot 10^{-5}$ & $\nu_4$; umbrella & $t_2$ & $a_1$ & $a_1$ \\ 
    $4$ & $1.892\cdot 10^{-4}$ & $\nu_3$; asymmetrical stretch & $t_2$ &
       $a_1$ & $a_1$ \\  
    $5$ & $4.894\cdot 10^{-5}$ & $\nu_2$; bending & $e$ & $e$ & $a_1$ \\
    $6$ & $4.894\cdot 10^{-5}$ & $\nu_2$; bending & $e$ & $e$ & $a_2$ \\
    $7$ & $3.599\cdot 10^{-5}$ & $\nu_4$; umbrella & $t_2$ & $e$ & $b_1$ \\ 
    $8$ & $3.599\cdot 10^{-5}$ & $\nu_4$; umbrella & $t_2$ & $e$ & $b_2$ \\  
    $9$ & $1.892\cdot 10^{-4}$ & $\nu_3$; asymmetrical stretch & $t_2$ &
       $e$ & $b_1$ \\  
    $10$ & $1.892\cdot 10^{-4}$& $\nu_3$; asymmetrical stretch & $t_2$ &
       $e$ & $b_2$ \\  
    \hline
  \end{tabular}
\end{table}
\begin{table}[H]
  \caption{$\alpha$ and $\beta$ values (in atomic units) of $V_{\rm
    surf}$ with one, two or three hydrogens pointing towards the surface
    (see Eqs.\  (\ref{vsurfone}), (\ref{vsurftwo}) and (\ref{vsurfthree})).} 
  \label{tab:alpha}
  \begin{tabular}{l r r r}
    \hline
    & one & two & three \\
    \hline
    $\alpha_1$ & $6.281\cdot 10^{-3}$  & $6.281\cdot 10^{-3}$  &
    $6.281\cdot 10^{-3}$ \\ 
    $\alpha_2$ & $1.256\cdot 10^{-2}$  & $7.252\cdot 10^{-3}$  &
    $4.187\cdot 10^{-3}$ \\
    $\alpha_3$ & $4.226\cdot 10^{-3}$  & $-7.931\cdot 10^{-3}$ &
    $-1.198\cdot 10^{-2}$ \\
    $\alpha_4$ & $-2.040\cdot 10^{-2}$ & $-7.445\cdot 10^{-3}$ &
    $-3.128\cdot 10^{-3}$ \\
    $\alpha_5$ &                       & $-1.026\cdot 10^{-2}$  &
    \\
    $\beta_{1}$ & &                          & $5.921\cdot 10^{-3}$ \\
    $\beta_{2}$ & &                          & $1.026\cdot 10^{-2}$ \\
    $\beta_{3}$ & & $6.079\cdot 10^{-3}$ & $2.866\cdot 10^{-3}$ \\
    $\beta_{4}$ & &                      & $4.963\cdot 10^{-3}$ \\
    $\beta_{5}$ & & $6.476\cdot 10^{-3}$ & $3.053\cdot 10^{-3}$ \\
    $\beta_{6}$ & &                      & $5.288\cdot 10^{-3}$ \\
    \hline
  \end{tabular}
\end{table}
\begin{table}[H]
  \caption{$\gamma$ values (in atomic units) of $V_{\rm Morse}$ with one
    and three hydrogens pointing towards the surface (see
    Eq.\  (\ref{Vmorse_d})).} 
  \label{tab:gammaone}
  \begin{tabular}{l l r}
    \hline
    one & three & value\\
    \hline
    $\gamma_{12},\gamma_{22}, \gamma_{32}, \gamma_{42}$ &
    $\gamma_{12}, \gamma_{22}, \gamma_{32}, \gamma_{42}$ &
    $1.079\cdot 10^{-2}$ \\ 
    $\gamma_{13}, -3\gamma_{23}, -3\gamma_{33}, -3\gamma_{43}$ & 
    $-\gamma_{13}, 3\gamma_{23}, 3\gamma_{33}, 3\gamma_{43}$ &
    $1.359\cdot 10^{-3}$ \\
    $\gamma_{14}, -3\gamma_{24}, -3\gamma_{34}, -3\gamma_{44}$ &
    $-\gamma_{14}, 3\gamma_{24}, 3\gamma_{34}, 3\gamma_{44}$  &
    $-1.966 \cdot 10^{-2}$ \\
    $\gamma_{17}, \gamma_{18}, \gamma_{19}, \gamma_{1,10}, \gamma_{28}, 
    \gamma_{2,10}$ & 
    $\gamma_{17}, \gamma_{18}, \gamma_{19}, \gamma_{1,10}, \gamma_{28},  
    \gamma_{2,10}$  & $0.0$ \\
    $\gamma_{27}, -2\gamma_{37}, -2\gamma_{47}$ &
    $-\gamma_{27}, 2\gamma_{37}, 2\gamma_{47}$ & $1.282 \cdot 10^{-3}$
    \\
    $\gamma_{38}, -\gamma_{48}$ & $\gamma_{38}, -\gamma_{48}$ &
    $-1.110 \cdot 10^{-3}$ \\ 
    $\gamma_{29}, -2\gamma_{39}, -2\gamma_{49}$ &
    $-\gamma_{29}, 2\gamma_{39}, 2\gamma_{49}$ & $-1.853 \cdot
    10^{-2}$ \\
    $\gamma_{3,10}, -\gamma_{4,10}$ & $-\gamma_{3,10}, \gamma_{4,10}$ &
    $1.605 \cdot 10^{-2}$ \\  
    \hline
  \end{tabular}
\end{table}

\begin{table}[H]
  \caption{$\gamma$ values (in atomic units) of $V_{\rm Morse}$ with two
    hydrogens pointing towards the surface (see Eq.\  (\ref{Vmorse_d})).}  
  \label{tab:gammatwo}
  \begin{tabular}{l  r}
    \hline
    two & value\\
    \hline
    $\gamma_{12}, \gamma_{22}, \gamma_{32}, \gamma_{42}$ & $1.079\cdot
    10^{-2}$ \\ 
    $\gamma_{13}, \gamma_{23}, -\gamma_{33}, -\gamma_{43}, \gamma_{17},
    -\gamma_{27}, \gamma_{37}, -\gamma_{47}, \gamma_{18}, -\gamma_{28},
    -\gamma_{38}, \gamma_{48}$ & $-7.849 \cdot 10^{-4}$  \\
    $\gamma_{14}, \gamma_{24}, -\gamma_{34}, -\gamma_{44}, \gamma_{19},
    -\gamma_{29}, \gamma_{39}, -\gamma_{49}, \gamma_{1,10},
    -\gamma_{2,10}, -\gamma_{3,10}, \gamma_{4,10}$ & $1.135 \cdot
    10^{-2}$ \\ 
    \hline
  \end{tabular}
\end{table}
\begin{table}[H]
  \caption{Excitation probabilities, at an initial translational energy
    of 96 kJ/mol and all initial vibrational states in the ground state,
    for the three different PESs in
    the $a_1$ modes of the C${}_{3v}$  and C${}_{2v}$ symmetry, with
    one, two or three hydrogens pointing towards the surface. 
    These modes are a $\nu_1(a_1)$ symmetrical stretch, a $\nu_2(e)$
    bending , a $\nu_3(t_2)$ asymmetrical stretch, and a $\nu_4(t_2)$
    umbrella in the T${}_d$ symmetry. The PESs are:
    An anharmonic intramolecular PES (Morse, see Eq.\ (\ref{Vmorse_d})),
    an intramolecular PES with weakening C--H bonds (weak, see
    Eq.\ (\ref{Vweak_d})), and an intramolecular PES with elongation of
    the C--H bonds (shift, see Eq.\ (\ref{Vshift_d})).}
  \label{tab:exprop}
  \begin{tabular}[t]{l l c c c c}
    \hline
      orientation & PES & $\nu_1(a_1)$ stretch & $\nu_2(e)$ bending &
      $\nu_3(t_2)$ stretch & $\nu_4(t_2)$ umbrella \\
    \hline
      one   & Morse & 0.023 &       & 0.054 & 0.030  \\
            & weak  & 0.135 &       & 0.308 & 0.075  \\
            & shift & 0.340 &       & 0.727 & 0.067 \\
    \hline
      two   & Morse & 0.010 & 0.108 & 0.009 & 0.104  \\
            & weak  & 0.067 & 0.135 & 0.073 & 0.107  \\
            & shift & 0.707 & 0.065 & 0.768 & 0.331 \\
    \hline
      three & Morse & 0.003 &       & 0.006 & 0.102  \\
            & weak  & 0.038 &       & 0.019 & 0.208  \\
            & shift & 0.819 &       & 0.674 & 0.214 \\
    \hline
  \end{tabular}
\end{table}
\begin{table}[H]
  \caption{Structure deformation, at an initial translational energy
    of 96 kJ/mol and all initial vibrational states in the ground state,
    with one, two, and three hydrogen 
    pointing towards the surface for three different PESs:
    An anharmonic intramolecular PES (Morse, see Eq.\ (\ref{Vmorse_d})),
    an intramolecular PES with weakening C--H bonds (weak, see
    Eq.\ (\ref{Vweak_d})), and an intramolecular PES with elongation of
    the C--H bonds (shift, see Eq.\ (\ref{Vshift_d})).
    See Fig.\ref{fig:orient} for the meaning $R_{\rm down}$, $R_{\rm
    up}$, $\alpha_{\rm down}$, $\alpha_{\rm up}$, and $\alpha_{\rm
    side}$. All bond distances are given in 
    {\AA} and all bond angles are given in degrees. The gas phase values
    correspond to the bond distance and angles far from the surface, and
    are the same for all three PES types.}
  \label{tab:strdef}
  \begin{tabular}[t]{l l c c c c c}
    \hline
      Orientation & PES & $R_{\rm down}$ & $R_{\rm up}$ & $\alpha_{\rm
      down}$ & $\alpha_{\rm up}$ &  $\alpha_{\rm side}$ \\
    \hline
    gas phase & all & 1.165 & & 109.5 & & \\
    \hline
      one   & Morse & 1.123 & 1.166 & 108.5 & 110.4 & \\
            & weak  & 1.051 & 1.167 & 108.2 & 110.7 & \\
            & shift & 1.397 & 1.158 & 109.9 & 109.0 & \\
    \hline
      two   & Morse & 1.157 & 1.165 & 116.7 & 108.9 & 107.8 \\
            & weak  & 1.137 & 1.163 & 117.6 & 109.0 & 107.5 \\
            & shift & 1.386 & 1.159 & 115.7 & 106.9 & 108.5 \\
    \hline
      three & Morse & 1.164 & 1.160 & 111.2 & 107.7 & \\
            & weak  & 1.167 & 1.155 & 112.4 & 106.3 & \\
            & shift & 1.389 & 1.144 & 111.2 & 107.7 & \\
    \hline
  \end{tabular}
\end{table}

%
% END of document
%
\end{document}